\documentclass[fleqn,usenatbib]{mnras}

\usepackage{newtxtext,newtxmath}

\usepackage[T1]{fontenc}
\usepackage{ae,aecompl}
\usepackage{flushend}
\usepackage{graphicx}	
\usepackage{amsmath}	
\usepackage{amssymb}	

\newcommand{\oau}{\rm{[\ion{O}{iii}]$\lambda$4363}}

\newcommand{\nii}{\rm{[\ion{N}{ii}]$\lambda$6584}}
\newcommand{\oiii}{\rm{[\ion{O}{iii}]$\lambda$5007}}
\newcommand{\oiiid}{\rm{[\ion{O}{iii}]$\lambda\lambda$4959,5007}}
\newcommand{\oii}{\rm{[\ion{O}{ii}]$\lambda$3727}}

\newcommand{\ha}{\rm H{$\alpha$}}
\newcommand{\hb}{\rm H{$\beta$}}

\newcommand{\heii}{\rm\ion{He}{ii}$\lambda$4686}
\newcommand{\gmass}{\log(M_*/M_{\sun})}

\title[Strong Emission Line Ratios Evolution]{What Drives the 
Redshift Evolution of Strong Emission Line Ratios?}

\author[Bian et al.]{Fuyan Bian$^{1,2}$,\thanks{E-mail: fbian@eso.org} Lisa J. Kewley$^{2,3}$,  Brent Groves$^{2,3}$, Michael A. Dopita$^{2,3}$\thanks{Deceased}
\\
$^{1}$European Southern Observatory, Alonso de C\'ordova 3107, Casilla 19001, Vitacura, Santiago 19,
Chile\\
$^{2}$Research School of Astronomy \& Astrophysics, Mt Stromlo Observatory, Australian National University, Canberra, ACT 2611, Australia\\
$^{3}$ARC Centre of Excellence for All Sky Astrophysics in 3 Dimensions (ASTRO 3D), Canberra, ACT 2611, Australia}

\date{Accepted XXX. Received YYY; in original form ZZZ}

\pubyear{2015}

\begin{document}
\label{firstpage}
\pagerange{\pageref{firstpage}--\pageref{lastpage}}
\maketitle

\begin{abstract}
We study the physical mechanisms that cause the offset between low-redshift and high-redshift galaxies on the {\oiii/\hb} versus {\nii/\ha} ``Baldwin, Phillips \& Terlevich'' (BPT) diagram using a sample of local analogues of high-redshift galaxies. These high-redshift analogue galaxies are selected from the Sloan Digital Sky Survey. Located in the same region on the BPT diagram as the ultra-violet selected galaxies at $z\sim2$, these high-redshift analogue galaxies provide an ideal local benchmark to study the offset between the local and high-redshift galaxies on the BPT diagram. We compare the nitrogen-to-oxygen ratio (N/O), the shape of the ionising radiation field, and ionisation parameters between the high-redshift analogues and a sample of local reference galaxies. The higher ionisation parameter in the high-redshift analogues is the dominant physical mechanism driving the BPT offset from low- to high-redshift, particularly at high {\nii/\ha}. Furthermore, the N/O ratio enhancement also plays a minor role to cause the BPT offset. However, the shape of the ionising radiation field is unlikely to cause the BPT offset because the high-redshift analogues have a similar hard ionising radiation field as local reference galaxies. This hard radiation field cannot be produced by the current standard stellar synthesis models. The stellar rotation and binarity may help solve the discrepancy. 
\end{abstract}




\section{Introduction}
Optical recombination and collisionally excited emission lines provide essential information to study the properties of the ionised interstellar medium (ISM) and ionising radiation field. The relative strength of these emission lines is regulated by the shape of the ionising radiation field, gas-phase chemical abundance, gas density, and ionisation parameter (the ratio of the ionising photon number density to the hydrogen number density). 

The  ``Baldwin, Phillips \& Terlevich'' \citep[BPT,][]{Baldwin:1981rr} diagrams are powerful tools to separate star-forming galaxies and active galactic nuclei \citep[AGNs, e.g.][]{Veilleux:1987aa}. Galaxies at $z\sim0$ are located in a well-defined star-forming sequence at on the {\oiii/\hb} versus {\nii/\ha} BPT diagram \citep[e.g.,][]{Stasinska:2006aa,Kewley:2006aa}. However, galaxies at $z\sim2$ do not share the same location of their local counterparts on the BPT diagram \citep[e.g.,][]{Erb:2006rt,Liu:2008aa,Hainline:2009aa,Bian:2010vn,Steidel:2014aa,Shapley:2015aa}. For a fixed {\nii/\ha} ratio, high-redshift galaxies tend to have a higher {\oiii/\hb} ratio, and vice versa. Understanding the dominant mechanism of cause this offset would provide valuable insight into how the stellar population, metal abundance, and star formation environment of galaxies evolve with cosmic time. 



The physical mechanism driving this offset is still under debate. Four competing interpretations have been proposed: (1) higher nitrogen-to-oxygen ratio \citep[N/O ratio, e.g.][]{Masters:2014aa,Masters:2016aa,Jones:2015aa,Shapley:2015aa,Kojima:2017aa}, (2) harder stellar radiation field \citep[e.g.,][]{Steidel:2014aa,Steidel:2016aa}, (3) higher ionisation parameter and/or electron density in high-redshift galaxies \citep[e.g.,][]{Liu:2008aa,Brinchmann:2008ab,Bian:2010vn,Kewley:2013ab,Kewley:2013aa,Kojima:2017aa,Dopita:2016aa}, and (4) selection effects \citep[e.g.,][]{Juneau:2014aa,Salim:2015aa}.

\citet{Bian:2016aa} found that selection effects cannot fully account for the offset between low- and high-redshift galaxies on the BPT diagram. Only less than 50\% of the galaxies selected from the Sloan Digital Sky Survey (SDSS) with the same emission-line luminosities or ultraviolet (UV) luminosities as high-redshift galaxies are located on the BPT star-forming sequence defined by $z\sim2$ UV-selected galaxies \citep{Steidel:2014aa}. Even if the selection effect can fully explain the offset, we still need to understand the physical mechanisms that cause it. Therefore, in this work, we focus on the first three of the interpretations of the observed BPT offset.

Strong optical emission lines, such as {\oii}, {\oiii}, {\hb}, {\ha},  and {\nii}, do not have the power to disentangle the above physical mechanisms, because these strong line ratios are highly degenerate with physical properties. For instance, a higher ionisation parameter, or a harder radiation field, or an increased pressure in the HII region can increase the {\oiii/\hb} ratio and shifts the low-redshift star-forming sequence to high-redshift star-forming sequence \citep[e.g.,][]{Kewley:2013ab}. Therefore, it is difficult to draw an exclusive conclusion based on these strong emission lines alone. In this work, we use a sample of local analogues of high-redshift galaxies to approach this issue. By stacking the spectra of these analogues and a sample of local reference galaxies, we are able to detect weak emission lines, including {\oau} and {\heii}, which usually cannot be detected in high-redshift galaxies. These weak lines provide powerful probes in disentangling the different physical mechanisms causing the BPT offset between the low- and high-redshift galaxies.

\begin{figure}
\begin{center}
\includegraphics[width=1.05\columnwidth]{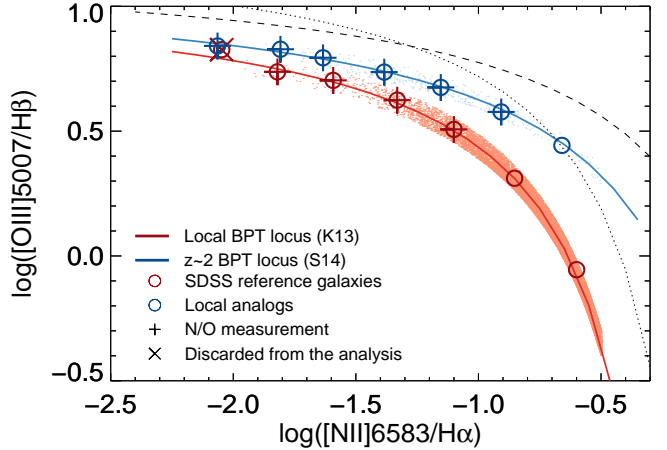}
\caption{BPT diagnostic diagram of local analogues of high-redshift galaxies and the SDSS reference galaxies. The small blue and red points represent the individual local analogues and the SDSS reference galaxies, respectively. The blue and red open circles represent the stacked spectra of the local analogues and SDSS reference galaxies, respectively. The open circles with plus symbols are the stacked spectra with reliable N/O ratio and oxygen abundance measurements using the direct $T_e$ method, and the open circle with `x' symbol denotes the spectrum discarded from our analysis. The blue solid line represents the star-forming BPT locus at $z\sim2$ adopted from \citet{Steidel:2014aa}, and the red solid line represents the local star-forming BPT locus adopted from \citet{Kewley:2013ab}.
The dotted and dashed lines represent the empirical \citep{Kauffmann:2003ij} and theoretical \citep{Kewley:2001aa} separations of star-forming galaxies
and AGNs, respectively.
\label{BPT}}
\end{center}
\end{figure}

\section{Method}
 \subsection{Sample Selection}\label{selection}
We select a sample of local analogues of high redshift galaxies and low-redshift reference galaxies from the SDSS galaxy survey. These galaxies are selected based on their locations on the [\ion{O}{iii}]/{\hb} versus [\ion{N}{ii}]/{\ha} BPT diagram (Figure~\ref{BPT}). The local analogues are selected within the $\pm0.04$~dex region of the $z\sim2$ star-forming sequence defined by equation~9 in \citet{Steidel:2014aa}. The SDSS reference galaxies are selected within the $\pm0.05$~dex region of the $z\sim0$ star-forming BPT sequence defined by equation~3 in \citet{Kewley:2013aa} (Figure~\ref{BPT}). We also use the [\ion{O}{iii}]/{\hb} versus [\ion{S}{ii}]/{\ha} and [\ion{O}{iii}]/{\hb} versus [\ion{O}{i}]/{\ha} BPT diagnostic diagrams to remove potential contamination from shock/AGN\citep{Bian:2018aa}. At last, a total of 443 galaxies are selected as local analogues of high-redshift galaxies, and a total of 22428 galaxies are selected as local reference galaxies. These local analogues of high-redshift galaxies share similar trend with star-forming galaxies at $z\sim2$ in all the diagnostic diagrams \citep[e.g.,][]{Steidel:2014aa,Steidel:2016aa, Shapley:2015aa, Sanders:2016aa,Strom:2017aa}. 

We summarise the physical properties of these high-redshift analogues as follows: The median stellar mass, SFR and sSFR of the sample of high-redshift analogous are $\gmass=8.8\substack{+0.06 \\ -0.02}$, $3.9\substack{+0.7 \\ -0.2}$~$M_{\sun}$~yr$^{-1}$, and $10.0\substack{+1.0 \\ -0.5}$~Gyr$^{-1}$, respectively. Their sSFR is consistent with that in $z\sim2$ star-forming galaxies with similar stellar mass \citep[e.g.,][]{Rodighiero:2011fk,Trainor:2016aa}, and these analogues follow the $M_*$-SFR relation at $z\sim2.3$ \citep{Sanders:2019aa}. It is worth noting that the local analogue selection is biased to the low stellar mass and low {\nii/\ha} ratio end comparing to the typical UV-selected or mass-selected galaxies at $z\sim2-3$, as it is discussed in \citet{Strom:2017aa}. Actually, these analogues more closely resemble the properties of Ly$\alpha$ emitters and low mass galaxies at $z\sim2-3$ \citep[e.g.][]{Trainor:2016aa}. Particularly, the ionisation parameters of these local analogues are in the range of $\log q=7.7 - 8.5$ ($\log U=-2.7$ to $\log U=-2.0$), which are well in agreement with those in $z\sim2.5$ Ly$\alpha$ emitters  \citep[e.g.][]{Nakajima:2014aa,Trainor:2016aa}. Furthermore, \citet{Sanders:2019aa} suggested that the metallicity calibrations derived from the direct $T_e$ metallicity in these local analogues reliably reproduce the average properties of $z>1$ galaxies with {\oau} auroral-line measurements, which are mostly low mass galaxies.

Therefore, it is a reasonable assumption that the physical mechanisms driving the offset between the local analogues and local star-forming sequence are the same as those determining the observed location of high-redshift star-forming galaxies on the BPT diagram. This type of local analogues provides an excellent local laboratory and benchmark to understand the physical mechanism(s) to drive the evolution of the [\ion{O}{iii}]/{\hb} versus [\ion{N}{ii}]/{\ha} BPT diagram between low- and high-redshift galaxies.

\subsection{Stacked Spectra}
By combining individual spectra we can detect the weak emission lines, including {\oau} and \ion{He}{ii}$\lambda$4686. The analogues of high-redshift galaxies and the SDSS reference galaxies have been divided into seven bins of 0.25 dex in the range $-2.25$ to $-0.25$ in their [\ion{N}{ii}]/{\ha} ratios. We refer the readers to Table 3 and Table 4 in \citet{Bian:2018aa} for detailed properties of galaxies in each of the [\ion{N}{ii}]/{\ha} bins.
We generate the stacked spectra in each [\ion{N}{ii}]/H$\alpha$ bin as follows. Each one-dimensional galaxy spectrum from the SDSS survey is deredshifted and dereddened based upon the Balmer decrement. The galaxy spectra are then normalized at the wavelength range between 4400{\AA} and 4450{\AA}. Finally, the spectra are combined by averaging the individual spectra in each [\ion{N}{ii}]/H$\alpha$ bin. 

We measure the line fluxes in the stacked spectra. The stellar continuum of each stacked spectrum is fitted using the STARLIGHT stellar population synthesis code \citep{Cid-Fernandes:2005aa} locally and then subtracted from the stacked spectra. Then the emission lines are fitted using Gaussian profiles to measure the line fluxes. Only two galaxies in the sample of local reference galaxy fall into the $-2.00<\log$([\ion{N}{ii}]/{\ha}$)<-2.25$ bin, and both of them are located above the local star-forming sequence, biasing the stacked spectrum toward the high-redshift star-forming sequence (Figure~\ref{BPT}). Therefore, this data point (the circle point with `x' symbol in Figure~\ref{BPT}) is discarded from further analysis. We detect the {\oau} emission line at {$\rm S/N>10$} in four of the stacked spectra of the SDSS reference galaxies and in six of the stacked spectra of the local analogues (the open circles with crosses in Figure~\ref{BPT}. The \ion{He}{ii}$\lambda4686$ lines are all well detected ($\rm S/N>10$) in all the stacking spectra. We refer readers to Section 3 in \citet{Bian:2018aa} for more detail for the procedure on stacked spectra, subtracting the stellar continuum, and emission-line fitting.

\section{Results}

\begin{figure}
\begin{center}
\includegraphics[width=1.0\columnwidth]{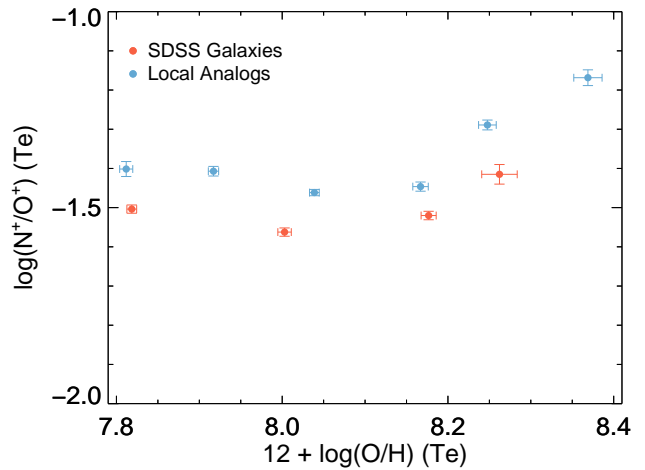}
\caption{N$^+$-to-O$^+$ ratio as a function of oxygen abundance in the local analogues of high-redshift galaxies (blue data points) and the SDSS reference galaxies (red data points). The N$^+$-to-O$^+$ ratio is a approximation to the N-to-O ratio \citep[e.g.,][]{Andrews:2013aa}. Both  N$^+$-to-O$^+$ ratio and oxygen abundance based on the direct $T_e$ method. The data points from left to right correspond to increasing {\nii/\ha} ratio direction.\label{NO-ratio}}
\end{center}
\end{figure}


\subsection{N-to-O ratios}
The stacked spectra enable us to study both the N/O ratio and the oxygen abundance in the local analogues and the local reference galaxies using the direct $T_e$ method. The $T_e$ method provides the most reliable ways to measure the N/O ratio and oxygen abundance. However, it is extremely challenging to measure the N/O ratio and oxygen abundance using the $T_e$ method in high-redshift galaxies \citep[e.g.,][]{Kojima:2017aa,Strom:2017aa,Sanders:2016ab}, because it requires a detection of the weak emission line {\oau}, which is more than 10 times weaker than {\hb} . However, such measurements are possible in this work, thanks to the high S/N in the stacked spectra of both the local analogues and the SDSS reference galaxies.

We compute the N/O ratio and oxygen abundance using the \citet{Izotov:2006ab} recipe. The electron temperature in the O$^{++}$ zone ($T_3$(O)) is derived using the {\oiiid}/{\oau} ratio. For the electron temperature in the O$^+$ zone ($T_2$(O)), the relation between $T_3$ and $T_2$ is adopted from \citet{Campbell:1986aa} as follows: $T_2=0.7T_3+3000$K. Our main conclusion is not sensitive to the $T_3$-$T_2$ relation used. We estimate O$^{++}$ and O$^+$ abundance based on the electron temperatures in the corresponding regions. The final oxygen abundance is derived by adding O$^{++}$ and O$^+$ abundance together. For the nitrogen abundance, we assume that the electron temperature in the N$^+$ zone is the same as that in the O$^+$ zone, $t_2$(N)=$t_2$(O) \citep[e.g.,][]{Kennicutt:2003aa}. We derive the N$^+$/O$^+$ ratio based on the $T_2$(N) and $T_2$(O), and the N$^+$/O$^+$ ratio is a good approximation to the N/O ratio \citep{Andrews:2013aa}. 

Figure~\ref{NO-ratio} shows the relation between N/O ratio and the oxygen abundance in the local analogues (blue data points) and the SDSS reference galaxies (red data points). The relation consists of two regimes: The primary nitrogen regime, where the N/O ratio is a constant at $12+\log(\rm{O/H})< 8.2$. The secondary nitrogen regime, where the N/O ratio increase with oxygen abundance at $12+\log(\rm{O/H})>8.2$ \citep[e.g.,][]{van-Zee:1998ab}.

The N/O ratios for the two samples of galaxies in the primary nitrogen regime is estimated by averaging the N/O ratios in the range $12+\log(\rm{O/H})< 8.2$.  We find that, in the primary N regime,  the N/O ratios in the local analogues are higher than those in the local reference galaxies by $\sim0.1$~dex at $-1.43 \pm 0.01$ and $-1.53\pm0.01$, respectively. The N/O offset between the local analogues and the SDSS reference galaxies in the secondary nitrogen regime is also at the similar level, but it is difficult to quantify the difference based on a few data points in this regime. The N/O offset is similar to that in green pea galaxies \citep[e.g.,][]{Amorin:2010aa}. Studies suggest that It requires the change of N/O by 0.2-0.4~dex to shift the BPT star-forming locus at $z\sim0$ to that at $z\sim2.3$ \citep[e.g.,][]{Masters:2016aa}. The N/O enhancement in the high-redshift analogues cannot fully explain the offset between the local analogues and the SDSS galaxies. \citet{Strom:2017aa} also found an $\sim0.1$~dex offset between $z\sim2$ star-forming galaxies and local galaxies using strong emission lines, and these authors also suggested the offset between $z\sim2$ and local star-forming galaxies on the BPT diagram can not fully account for the N/O enhancement.

It is worth noting that our main conclusion will not change, if we adopt different recipes to estimate $T_3$(O) \citep[e.g,][]{Pagel:1992aa},  different $T_3$-$T_2$ relations \citep[e.g.,][]{Lopez-Sanchez:2012aa}, and different 
$t_2$(N)-$t_2$(O) relations \citep[e.g.,][]{Berg:2015aa}.

\begin{figure*}
\begin{center}
\includegraphics[width=2.2\columnwidth]{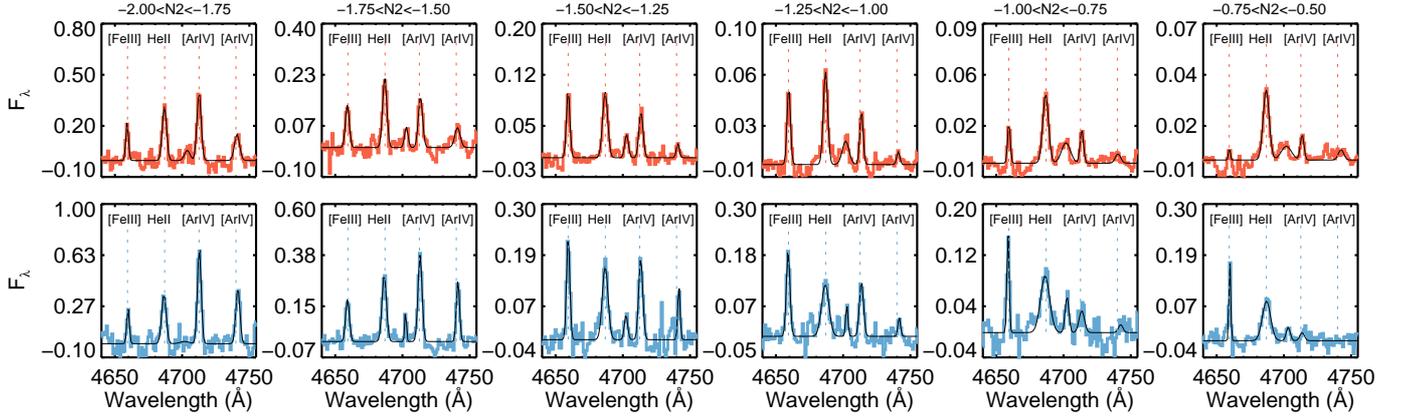}
\caption{The {\heii} spectra of the SDSS reference galaxies (top panel) and the local analogues of high-redshift galaxies (bottom panel) in different {\nii/\ha} bins. The $\log${\nii/\ha} (N2) increases from left to right. The most right panel is for the bin of $-0.75<$$\log$({\nii/\ha})$<-0.50$, and the most left panel is for the bin of $-2.00<$$\log$({\nii/\ha})$<-1.75$
\label{spec_he}}
\end{center}
\end{figure*}

\begin{figure}
\begin{center}
\includegraphics[width=1.0\columnwidth]{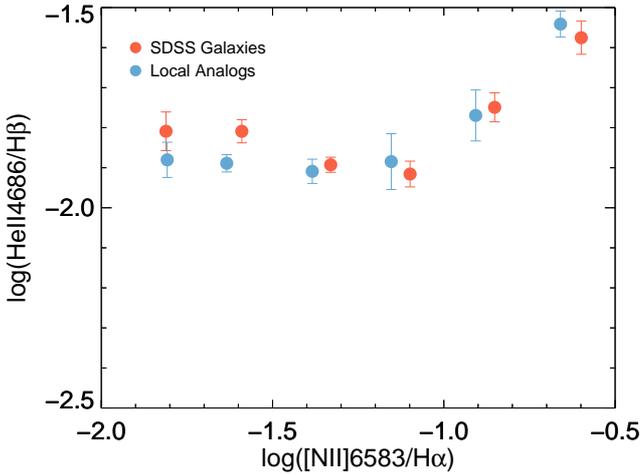}
\caption{{\heii/\hb}\label{bpt_he} vs. {\nii/\ha} diagram of the local analogues of high-redshift galaxies (blue data points) and the SDSS reference galaxies (red data points). The black lines are the best-fit Gaussian Models of the emission lines}
\end{center}
\end{figure}

\subsection{Spectral Hardness}
The spectral hardness can be studied using the {\heii} emission line, because the ionisation potential of He$^{++}$  is 54.4eV, thus the {\heii} emission line is very sensitive to the hard ionising radiation field. Figure~\ref{spec_he} shows the stacked spectra around {\heii} in the high-redshift analogues and local reference galaxies in six [\ion{N}{ii}]/{\ha} bins in the range of $-2.00<$$\log$({\nii/\ha})$<-0.50$. We fit the {\heii} line with other nearby emission lines, including [\ion{Fe}{iii}] and [\ion{Ar}{iv}] simultaneously. We firmly (${\rm S/N}>10$) detect the {\heii} in all stacked spectra. The {\heii} line width is $\sigma=100-200$~km~s$^{-1}$ without obvious underlying broad component (Figure~\ref{spec_he}). The emission line width of {\rm \ion{He}{II}} is consistent with that of H$\beta$ in the N2 bins with N2 $ < -1.25$. However, when N2 $ >-1.25$, the {\rm \ion{He}{II}} line width is systematically larger than H$\beta$ line width by about 50 km~s$^{-1}$ in velocity dispersion. The {\rm \ion{He}{II}} line width suggests that the {\heii} emission is mostly raised from nebular emission at N2 $ < -1.25$, and at N2 $ >-1.25$ the atmospheric emission in Wolf-Rayet stars also starts to contribute to the {\heii} emission, however, the {\heii} emission in Wolf-Rayet is not the dominate component of the {\heii} emission in these galaxies due to relatively narrow {\heii} line width, a few hundred km~s$^{-1}$ rather than a few thousand km~s$^{-1}$.

Figure~\ref{bpt_he} shows the {\heii}/{\hb} ratio as a measure of spectral hardness in the local analogues and the SDSS reference galaxies in different [\ion{N}{ii}]/{\ha} bins. For $\log$([\ion{N}{ii}]/{\ha}$) <-1.2$, the {\heii}/{\hb} slightly decreases with increasing [\ion{N}{ii}]/{\ha} ratio, but then slightly increases with further increase in the [\ion{N}{ii}]/{\ha} ratio. This increase is presumably correlated with a greater fraction of W-R stars at higher metallicity. This trend is generally consistent with that found in the individual galaxies with {\heii} detection \citep{Shirazi:2012aa}.
We notice that the {\heii}/{\hb} ratio in the stacked is slightly (<0.5\%) higher than the {\heii}/{\hb} ratio measured in the individual SDSS galaxies \citep{Shirazi:2012aa}, especially at the high [\ion{N}{ii}]/{\ha} end. This could be due to the high S/N spectra of the stack spectra, which allows us better fitting the flux from underlying weak broad component of the \heii emission line.

We find that the {\heii}/{\hb} values of the local analogues and the SDSS reference galaxies are consistent with each other in all the [\ion{N}{ii}]/{\ha} bins, suggesting that the local analogues of high-redshift galaxies and the SDSS reference galaxies share a similar shape of the ionising radiation field.

\begin{figure}
\begin{center}
\includegraphics[width=1.0\columnwidth]{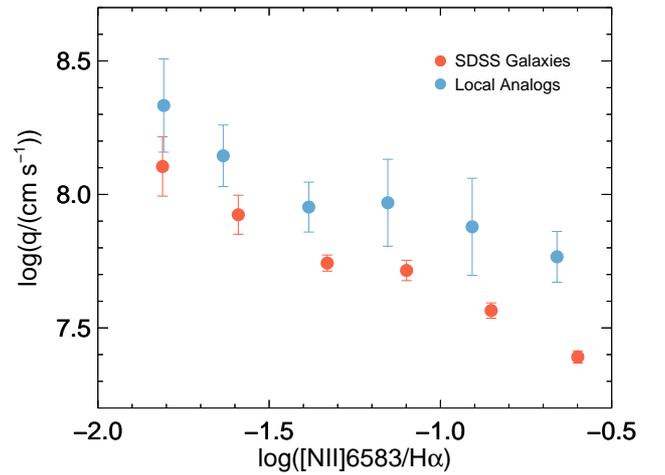}
\caption{ionisation parameter vs. $\log$(\nii/{\ha}) in the local analogues of high-redshift galaxies (blue data points) and the SDSS reference galaxies (red data points).
\label{logq}}
\end{center}
\end{figure}

\subsection{Ionisation Parameter}
The ionisation parameter is estimated for each stacked spectrum using the $O32 = $([\ion{O}{iii}]$\lambda\lambda$4959,5007/[\ion{O}{ii}]$\lambda$3727) and $R23 = $[([\ion{O}{ii}]$\lambda$3727+[\ion{O}{iii}]$\lambda\lambda$4959,5007)/H$\beta$] ratios. We adopt the \citet{Kobulnicky:2004aa} recipe to compute the metallicity and the ionisation parameter iteratively until the metallicity converges. We refer readers to \citet{Kobulnicky:2004aa} and \citet{kewley:2008aa} for more details. 

Figure~\ref{logq} shows the ionisation parameter as a function $\log$([\ion{N}{ii}]/{\ha}) in the local analogues and the local reference galaxies. In both galaxy samples, the ionisation parameters decrease with [\ion{N}{ii}]/{\ha}. The local analogues have significantly higher ionisation parameters than the local reference galaxies in all $\log$([\ion{N}{ii}]/{\ha}) bins. The mean difference of ionisation parameters between the local analogues and local reference galaxies is $0.27\pm0.05$~dex. The difference of the ionisation parameter increases with [\ion{N}{ii}]/{\ha}. In the highest [\ion{N}{ii}]/{\ha} bin, the ionisation parameter in the local analogues is $0.38\pm0.09$~dex higher than that in the local reference galaxies, and the difference of ionisation parameter becomes $0.23\pm0.20$~dex in the lowest [\ion{N}{ii}]/{\ha} bin. The ionisation parameter plays a major role to drive the BPT evolution with redshift, particularly at he high metallicity end.

By applying photo-ionization models to cosmological zoom-in simulations, \citet{Hirschmann:2017aa} found a similar result that evolution of the \oiii/{\hb} vs. {\nii}/{\ha} emission ratios is mainly driven by ionisation parameters. In their models, the ionisation parameter is directly regulated by the SFR. This is in agreement with the observational studies that have shown the ionisation parameter increases with SFR and specific SFR (SFR/$M_*$) \citep[e.g.,][]{Kaasinen:2018aa}. Our high-redshift analogues also have high specific SFR \citep{Bian:2016aa} and follow the $z\sim2.3$ $M_*$-SFR relation \citep{Sanders:2019aa}. Thus the high (specific) SFR is the main physical origin to drive the high ionisation parameter and the evolution of the \oiii/{\hb} vs. {\nii}/{\ha} emission ratios.

\section{Discussion}
\subsection{Origin of HeII Emission}
The {\heii} emission strength is a few percent of {\hb} in the stacked spectra in both the high-redshift analogues and local reference galaxies. This strength is similar to that in the individual galaxies that were selected from the SDSS survey based on their strong {\heii} emissions \citep{Shirazi:2012aa}. Such strong {\heii} emission is proposed to arise from the atmosphere of Wolf-Rayet stars and/or nebular emission due to the hard radiation field emitted from the Wolf-Rayet stars. However, the lifetime of Wolf-Rayet stars is short, on the order of $\sim2$ million years. Given this, Wolf-Rayet galaxies should only be a short transitional phase of galaxies,  and only a small fraction of galaxies should contribute to the {\heii} emission when stacking the spectra. Therefore, the {\heii}/{\hb} ratio in the stacked spectra should be significantly smaller than that in the individual Wolf-Rayet galaxies. The similar strength of the {\heii} in the stacked spectra and individual Wolf-Rayet galaxies suggests that the {\heii} emission commonly exists in a majority of these populations of galaxies and is unlikely due to the hard radiation field of Wolf-Rayet stars. Furthermore, the {\heii} emissions in the stacked spectra are narrow with $\sigma=100-200$~km~s$^{-1}$ with no obvious broad component, associated with strong Wolf-Rayet stars \citep[e.g.,][]{Miralles-Caballero:2016aa}.

We consider the following potential sources to produce the {\heii} emissions. 
\begin{enumerate}
\item A hard radiation field from AGNs or shocks can excite the {\heii} emission. 
Shocks and AGN move galaxies from the star-forming sequence to the AGN/shock region in the BPT diagnostic diagrams. In Section~\ref{selection},  we selected galaxies in all three BPT diagrams to minimise the AGN/shock contamination. In particular, the SDSS reference galaxy sample is located right on the local star-forming sequence on the BPT diagram and well separated from the AGN and shock regions. Therefore,  shocks and AGNs contribution should be negligible in both the local analogues and the SDSS reference galaxies. 

\item X-ray binaries can provide a hard radiation field to produce {\heii} emission. \citet{Kaaret:2004aa} studied the nebular close to an ultra-luminous X-ray source in Holmberg II and establish a relation between the nebular {\heii} and X-ray luminosity, $L_{\rm He II}\sim 10^{-3} L_{\rm X-ray}$, using CLOUDY photo-ionisation models. X-ray luminosity in star-forming galaxies is dominated by X-ray emission from the high mass X-ray binaries. In this situation, X-ray luminosity can be used as a star formation rate indicator \citep[e.g.,][]{Ranalli:2003aa}. We connect the X-ray luminosity to H$\beta$ luminosity, $L_{\rm X-ray}\sim10^{-2}L_{\rm H\beta}$, which leads to $L_{\rm He II} = 10^{-6} L_{\rm H\beta}$. Therefore, we estimate  the {\heii} emission from the X-ray binaries to be 4 orders of magnitude smaller than that in the stacked spectra.

\item The hard radiation field is due to binary stars and/or fast rotating stars. Integrating the binary stars and stellar rotation in the stellar synthesis models could significantly change the stellar evolution track of massive stars and increase their lifetime \citep[e.g.,][]{Eldridge:2009aa}. Therefore, binary evolution models result in the presence of massive stars (e.g., Wolf-Rayet stars) for a wider range of ages, which likely increases the contribution from stellar {\heii} emission in the stacked spectra. In the stellar synthesis models taking binary evolution into account (e.g., BPASS), the time scale over which stars can produce He$^{++}$ ionising photons is on the order of 100 Myr, which is two orders of magnitude higher than that derived from the stellar synthesis models that do not take binary evolution into account \citep[e.g.,][]{Wofford:2016aa}. The flux ratio of {\heii} to H$\beta$ has been measured in individual \ion{H}{ii} regions using deep optical spectra to be on the order of one percent level \citep[e.g.,][]{Berg:2015aa}. Given this, we believe the {\heii} emission is mostly likely due to the hard radiation field from massive stars, particularly when taking the binary evolution models into account. Such a scenario has also been used to explain the nebular emission lines requiring similar high excitation energies, such as \ion{He}{ii}1640, \ion{C}{iv}1549, \ion{C}{iii}]1909, detected in both low and high-redshift galaxies \citep[e.g.,][]{Stark:2014aa,Stark:2015aa,Senchyna:2017aa}.
\end{enumerate}


\section{Conclusion}
We study the physical origins of the offset between the low- and high-redshift galaxies on the {\oiii/\hb} versus {\nii/\ha} BPT diagram. We select a sample of local analogues of high-redshift galaxies and the local reference galaxies from the SDSS survey based on their location of the BPT diagram. The SDSS reference galaxy located on the local star-forming BPT sequence, and the local analogue galaxies located on the high-redshift star-forming BPT sequence. The local analogues well resemble the physical properties of high-redshift star-forming galaxies, provide an ideal laboratory to study what causes the evolution of star-forming BPT sequence from high- to low-redshift. 

Using the {\oau} and {\heii} weak emission lines in the stacked spectra, we found (1) the N/O ratios in the local analogues of high-redshift galaxies are $\sim0.1$~dex higher than those in the SDSS reference galaxies, which play a minor role (25\%-50\%) to drive the offset on the BPT diagram; (2) the ionisation parameters in the local analogues are 0.3-0.4~dex higher than those in the SDSS reference galaxies, which are the major cause (50\%-75\%) of the offset on the BPT diagram; (3) the local analogues and the SDSS reference have similar hard ionising radiation field. This hard radiation is mostly likely to originate from stars with significant rotation and/or in binary systems.





\section*{Acknowledgements}
F.B. thanks C. Steidel, A. Shapley, M. Pettini, X. Fan, H. Feng, and P. Martini for useful discussions of the work. We would like to thank the anonymous referee for providing constructive comments and help in improving the manuscript. LK gratefully acknowledges support from an ARC Laureate Fellowship
(FL150100113). BG gratefully acknowledges the support of the Australian Research Council as the recipient of a Future Fellowship (FT140101202).
MD acknowledges the support of the Australian Research Council (ARC) through Discovery project DP16010363. Parts of this research were conducted by the Australian Research Council Centre of Excellence for All Sky Astrophysics in 3 Dimensions (ASTRO 3D), through project number CE170100013.

\bibliography{paper}

\bsp	
\label{lastpage}
\end{document}